\documentclass[journal=nalefd,manuscript=letter,layout=traditional,citeautoscript]{achemso}

\usepackage{xr}
\makeatletter
\newcommand*{\addFileDependency}[1]{
  \typeout{(#1)}
  \@addtofilelist{#1}
  \IfFileExists{#1}{}{\typeout{No file #1.}}
}
\makeatother

\newcommand*{\myexternaldocument}[1]{
    \externaldocument[S]{#1}
    \addFileDependency{#1.tex}
    \addFileDependency{#1.aux}
}

\myexternaldocument{SI} 

\usepackage{array}
\usepackage{amsmath}
\usepackage{hyperref}
\usepackage{cleveref}
\usepackage{enumitem}
\usepackage{float}
\usepackage{gensymb}
\usepackage{multirow}
\usepackage{graphicx}
\usepackage{graphics}
\usepackage{dcolumn}
\usepackage{siunitx}
\usepackage{xr-hyper}
\usepackage{bm}

\usepackage{xargs}                      
\usepackage[pdftex,dvipsnames]{xcolor}  
\usepackage[colorinlistoftodos,prependcaption,textsize=tiny]{todonotes}

\newcommandx{\unsure}[2][1=]{\todo[linecolor=red,backgroundcolor=red!25,bordercolor=red,#1]{#2}}
\newcommandx{\change}[2][1=]{\todo[linecolor=blue,backgroundcolor=blue!25,bordercolor=blue,#1]{#2}}
\newcommandx{\info}[2][1=]{\todo[linecolor=OliveGreen,backgroundcolor=OliveGreen!25,bordercolor=OliveGreen,#1]{#2}}
\newcommandx{\improvement}[2][1=]{\todo[linecolor=Plum,backgroundcolor=Plum!25,bordercolor=Plum,#1]{#2}}
\newcommandx{\thiswillnotshow}[2][1=]{\todo[disable,#1]{#2}}

\title{Identification and Structural Characterization of Twisted Atomically Thin Bilayer Materials by Deep Learning}

\author{Haitao Yang}
\altaffiliation{These authors contributed equally to this work}
\affiliation{School of Advanced Materials and Nanotechnology, Xidian University, Xi’an 710071, China}

\author{Ruiqi Hu}
\altaffiliation{These authors contributed equally to this work}

\affiliation{Department of Materials Science and Engineering, University of Delaware, Newark, DE 19716, USA}

\author{Heng Wu}
\altaffiliation{These authors contributed equally to this work}
\affiliation{State Key Laboratory of Superlattices and Microstructures, Institute of Semiconductors, Chinese Academy of Sciences, Beijing
 100083, China}
 
\author{Xiaolong He}
\altaffiliation{These authors contributed equally to this work}
\affiliation{School of Advanced Materials and Nanotechnology, Xidian University, Xi’an 710071, China}

\author{Yan Zhou}
\email{yan.zhou@bristol.ac.uk}
\affiliation{State Key Laboratory of Superlattices and Microstructures, Institute of Semiconductors, Chinese Academy of Sciences, Beijing 100083, China}
\affiliation{Phonon Engineering Research Center of Jiangsu Province,  School of Physics and Technology, Nanjing Normal University, Nanjing 210023, China. }

\author{Yizhe Xue}
\affiliation{School of Advanced Materials and Nanotechnology, Xidian University, Xi’an 710071, China}

\author{Kexin He}
\affiliation{ School of Advanced Materials and Nanotechnology, Xidian University, Xi’an 710071, China}

\author{Wenshuai Hu}
\affiliation{School of Advanced Materials and Nanotechnology, Xidian University, Xi’an 710071, China}

\author{Haosen Chen}
\affiliation{School of Advanced Materials and Nanotechnology, Xidian University, Xi’an 710071, China}

\author{Mingming Gong}
\affiliation{School of Materials Science and Engineering, Northwestern Polytechnical University, Xi’an 710072, China}

\author{Xin Zhang}
\affiliation{State Key Laboratory of Superlattices and Microstructures, Institute of Semiconductors, Chinese Academy of Sciences, Beijing 100083, China}

\author{Ping-Heng Tan}
\email{phtan@semi.ac.cn}
\affiliation{State Key Laboratory of Superlattices and Microstructures, Institute of Semiconductors, Chinese Academy of Sciences, Beijing
100083, China}

\author{Eduardo R. Hern\'{a}ndez}
 \email{Eduardo.Hernandez@csic.es}
\affiliation{Instituto de Ciencia de Materiales de Madrid (ICMM-CSIC), 28049 Madrid, Spain}

\author{Yong Xie}
\email{yxie@xidian.edu.cn}
\affiliation{ School of Advanced Materials and Nanotechnology, Xidian University, Xi’an 710071, China}
\alsoaffiliation{Instituto de Ciencia de Materiales de Madrid (ICMM-CSIC), 28049 Madrid, Spain}

\begin{document}

\maketitle

\clearpage
\begin{abstract}
Two-dimensional materials are expected to play an important role in next-generation electronics and optoelectronic devices. Recently, twisted bilayer graphene and transition metal dichalcogenides have attracted significant attention due to their unique physical properties and potential applications. In this study we describe the use of optical microscopy to collect the color space of chemical vapor deposition (CVD) molybdenum disulfide ($\mbox{MoS}_2$), and the application of a semantic segmentation convolutional neural network (CNN) to accurately and rapidly identify thicknesses of $\mbox{MoS}_2$ flakes. A second CNN model is trained to provide precise predictions on the twist angle of CVD-grown bilayer flakes. This model harnessed a dataset comprising over 10,000 synthetic images, encompassing geometries spanning from hexagonal to triangular shapes. Subsequent validation of the deep learning predictions on twist angles was executed through the second harmonic generation and Raman spectroscopy. Our results introduce a scalable methodology for automated inspection of twisted atomically thin CVD-grown bilayer.

\end{abstract}


\begin{tocentry}
\includegraphics[width=1\linewidth]{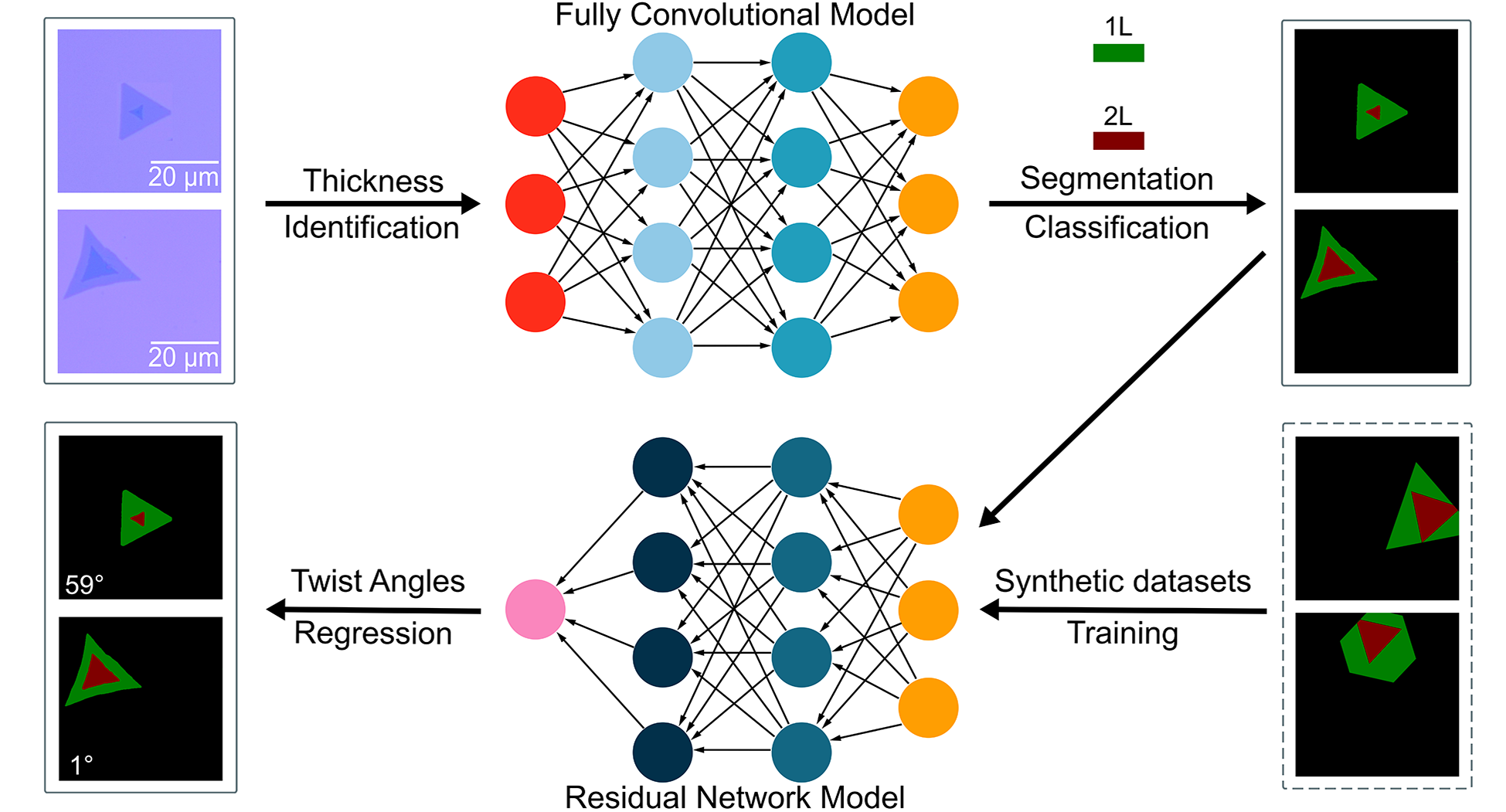}
\end{tocentry}

\clearpage

Inspired by magic-angle graphene\cite{Cao2018magicangle,Cao2018correlated}, twisted bilayer graphene and transition metal dichalcogenides~(TMDs) have emerged as a promising platform for the study of moir\'{e} physics, encompassing a range of phenomena such as Hubbard physics~\cite{tang2020simulation,wu2018hubbard}, superconductivity~\cite{wang2020correlated}, or valley polarization~\cite{devakul2021magic}. The twist angle in TMDs bilayers can significantly alter their correlated electronic phases and their optical properties~\cite{Lau2022review}. For example, in the twist angle ($2^\circ \leq \theta < 6^\circ$), low-frequency interlayer shear and layer breathing modes exhibit rapid change with the twist angle~$\theta$~\cite{quan2021phonon}. Additionally, the formation of the moir\'{e} Brillouin zone introduces new energy subbands in twisted $\mbox{MoS}_2$ bilayers with twist angles close to $0^\circ$ or $60^\circ$~\cite{marcellina2021evidence}, or high-lying excitons in bilayer $\mbox{WSe}_2$, which can be tuned over 235~meV by enforcing different twist angles in the range $0^\circ$ to $60^\circ$~\cite{lin2021twist}. The electric field control of the 2H bilayer $\mbox{MoS}_2$ interlayer exciton at room temperature is possible due to the out-of-plane electric dipole~\cite{peimyoo2021electrical}.

Chemical vapor deposition~(CVD) can be used to fabricate bilayer graphene\cite{liu2021graphene} and bilayer TMDs with different twist angles~\cite{Xie2017,wang2017nacl, paradisanos2020, zhang2019transition,WANG2020144371}, {\em i.e.\/} different stacking arrangements between the two layers. Typically, only $0^\circ$ (AA stacking, or 3R) or $60^\circ$ (AB stacking, or 2H) arrangements are possible for the second layer on bilayer TMDs, as these are energetically the more favourable~\cite{zhang2019transition,dumcenco2015large}. Reflectivity spectra have shown an A and B exciton energy difference of 49~meV between the 2H and 3R bilayer $\mbox{MoS}_2$~\cite{paradisanos2020}. Reverse-flow chemical vapor epitaxy provided a way to controllable grow high-quality bilayer TMD single crystals with different growth temperatures~\cite{zhang2019transition}.

Second Harmonic Generation (SHG) spectroscopy is frequently employed to characterize the twist angles of TMDs~\cite{yin2014, PhysRevB.87.161403}. This reliable method determines the orientation of exfoliated flakes and subsequently enables the stacking of homo- or hetero- layers using a dry transfer technique~\cite{castellanos2014deterministic,wang2013one,cheng2015kinetic}. In addition to SHG, differential reflectance spectroscopy can also be used for characterizing TMDs, specifically by identifying twist angles through the transition of interlayer excitons~\cite{Carrascoso2020, paradisanos2020}. Raman spectroscopy, particularly low wavenumber Raman spectroscopy~\cite{Tan-2012-NM,Zhang-2013-PRB}, is another common method used for this purpose~\cite{Zande2014Twist,Lin2018Twisted, zhang2019transition,quan2021phonon,wu2023analyzing}. For cases where atomic precision is required, Transmission Electron Microscopy (TEM) can be employed to determine the twist angles of graphene~\cite{Yoo2019bilayergraphene}. Further augmenting the atomic precision of characterization techniques, the Scanning Tunneling Microscope (STM) offers local probing of twist angles with atomic level resolution~\cite{Liu2015Bilayer}. Although accurate and reliable, these experimental techniques are costly, require specialised equipment and are time consuming. It is therefore desirable to develop alternative structural characterisation techniques that are cost effective, fast and easily implemented without compromising accuracy and reliability.

Although the thickness is normally confirmed by atomic force microscopy (AFM), Raman, etc \cite{li2012bulk}, optical contrast is frequently adopted by experience researcher in terms of the speed and simplicity based on the optical contrast \cite{castellanos2010optical}. Recent developments in artificial intelligence~(AI) have led to the adoption of new techniques for processing microscopy image datasets of layer thicknesses, edges, dimensions, etc.~\cite{frisenda2018robotic, masubuchi2018autonomous, lin2018intelligent}. For example, autonomous robotic search and stacking of graphene flakes were proposed to detect up to 400 monolayer graphene samples in one hour~\cite{frisenda2018robotic, masubuchi2018autonomous}. Unsupervised Machine Learning~(ML) and Deep Learning~(DL) techniques have been used to classify two-dimensional~(2D) materials into different categories~\cite{masubuchi2019classifying,han2020deep,masubuchi2020deep,sterbentz2021universal}. In combination with an automatic optical microscope stage, the desired~2D materials can be searched automatically~\cite{masubuchi2020deep}. However, to the best of our knowledge, no automatic procedure for determining the twist angle in bilayer atomically thin materials (e.g. TMDs and graphene) has been described up to now. Presumably, the DL model may lack sufficient accuracy due to the absence of adequate experimental data for training.

In this work, a systematic methodology is reported that demonstrates the potential of DL and image processing tools for determining the twist angles in CVD-grown bilayer TMDs and graphene. Specifically, we trained four different DL algorithms to identify the thickness of the CVD-grown flakes. The twist angle in individual bilayer TMDs flakes can be estimated by means of image processing tools, such as implemented in OpenCV~\cite{OpenCV}. This procedure, illustrated below, is nevertheless slow and not effective for large-scale sample analysis. To circumvent this problem, we have developed a second DL model to characterize the twist angle of individual flakes in an efficient way. All codes and datasets are open access and freely available, provided with user-friendly instructions. Our work aims to provide new tools designed to facilitate and make more effectively structural characterization of CVD-grown twisted TMDs samples, with extended applicability to CVD-grown graphene and hexagonal boron nitride (h-BN) and other CVD-grown two-dimensional materials.

\begin{figure}[ht]
    \centering
    \includegraphics[width=\linewidth]{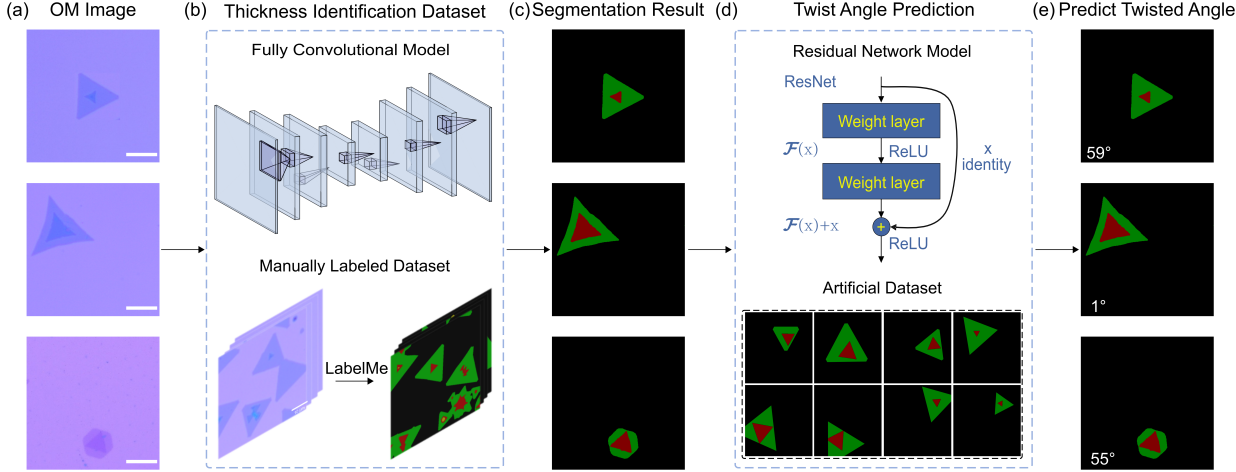}
    \caption{{Identification and analysis of optical micrographs of CVD-grown bilayer atomically thin materials (e.g. TMDs) using deep learning. Images of the TMDs are shown as the example.} (a) Original optical micrographs (OMs) captured by an Optical Microscope. (b) Images processed and labeled using LabelMe, followed by training via convolutional neural network (CNN) employing various classification methods. (c) Typical outcomes of the TMD thickness derived from the processing in step (b). (d) The regression CNN model trained by the artificially generated dataset for twist-angle prediction. (e) Twist angles predicted by using the CNN model from (d) are shown at the left corner at each image. Scale bar: $10\,\mu\text{m}$.}
    \label{fig_1}
\end{figure}

In our study, optical micrographs of CVD-grown bilayer atomically thin materials are captured as shown in Figure~\ref{fig_1} (for $\mbox{MoS}_2$) and supplementary for graphene \cite{liu2021graphene} (Figure S15).  These images are then processed and utilized to train a convolutional neural network (CNN) for the identification of flake thickness. Subsequently, a different CNN model, developed using a synthetic dataset, is used to predict the twist angles of the flakes, with these predictions displayed on the corresponding images.  The workflow diagram of the process to identify the twist bilayer of TMDs is visually shown in the supplementary Figure S1.

\begin{figure}[H]
\centering
\includegraphics[width=\linewidth]{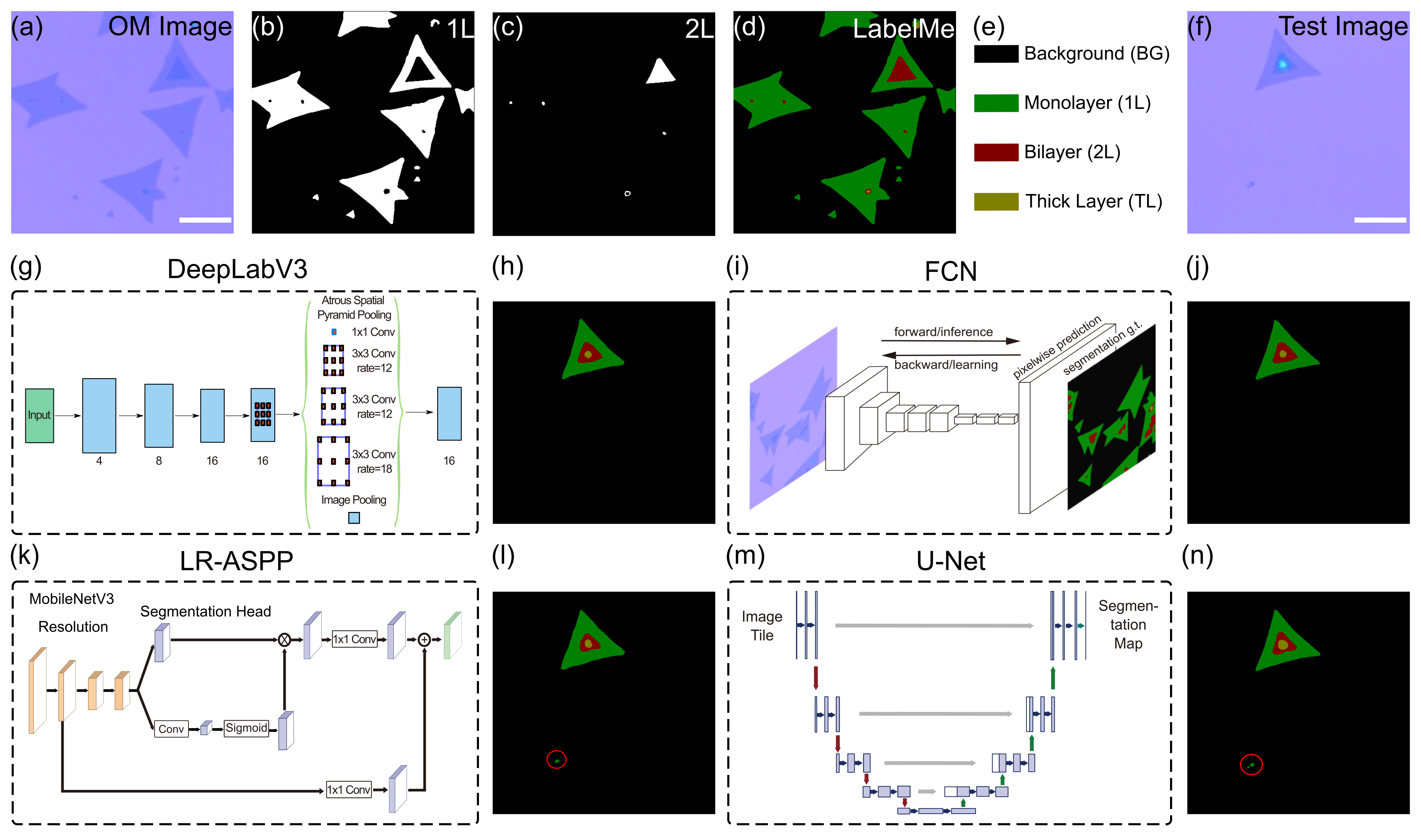}
\captionsetup{width=\textwidth}
\caption{\label{fig_2}
Segmentation techniques for classifying thickness in atomically thin CVD-grown bilayer flakes. (a) Optical micrographs of a bilayer $\mbox{MoS}_2$ flake. (b-c)  Detail the labeling for single and bilayer regions. (d) Final labeled result identifying thickness variations.(e) Explains the color labels, correlating colors to specific thickness levels. (f) Optical micrographs of another bilayer flake using as a test image.
 (g-n) Four CNN models and segmentation results employing \emph{DeepLabV3}, \emph{FCN}, \emph{LR-ASPP}, and \emph{U-Net} models respectively. Notably, the \emph{U-Net} model excels in recognizing the contours of imperfect triangular flakes.Scale bar: $10\,\mu\text{m}$.
}
\end{figure}

We utilized deep learning techniques to determine the thickness of atomically thin flakes, harnessing a supervised neural network trained on manually labeled images. These atomically thin flakes were initially distinguished by optical contrast and subsequently verified through AFM and Raman spectroscopy~\cite{Xie2017,wang2017nacl}, detailed in Supplementary Section 1.1.

Figure~\ref{fig_2} showcases the adeptness of our segmentation deep learning (DL) models in classifying flakes into monolayers (1L), bilayers (2L), and thicker layers (TL) with remarkable precision, as further elaborated in Figure 2 and Figure S7.  Figure~\ref{fig_2}(a) shows an original unprocessed microscopy image. Figures~\ref{fig_2}(b-e) illustrate the manual labeling of background, monolayer, bilayer, and thick layers.

In our evaluation, four DL models were rigorously tested: DeepLabV3 \cite{chen2017rethinking}, Fully Convolutional Network (FCN) \cite{long2015fully}, MobileNetV3 \cite{howard2019searching}, and U-Net \cite{ronneberger2015u}, each employing distinct segmentation strategies and architectures. For instance, DeepLabV3 integrates a backbone architecture for feature extraction with advanced techniques like dilated convolutions and spatial pyramid pooling. The FCN model, tailored for semantic segmentation, ensures the output size matches the input. Meanwhile, LR-ASPP, a variant of DeepLabV3's ASPP, aims for efficiency in mobile and edge computing. Notably, U-Net, recognized for its effectiveness in biomedical image segmentation, features a distinctive U-shaped architecture. Our implementation utilized Python 3.8 and Pytorch 1.10 \cite{torchvision2016}, with U-Net emerging as the most adept in capturing the nuances of shape, particularly evident in cases of distorted triangular overlays. Compared to the original micrographs, U-Net's segmentation reveals enhanced detail in both monolayer and bilayer configurations (Figure~\ref{fig_2}(n)), affirming its selection as the primary model for our study. For a comprehensive understanding of the labeling process and additional model comparisons, readers are directed to the Supplementary Information (Supplementary 2.1).

Once the pixels in an experimental image have been classified, either by a human or by any of the trained DL models described in Sec.\,I of SI, and the resulting flakes color-coded according to their thickness, it is possible to determine the twist angle formed between an underlying single-layer and an overlying second layer using appropriate image analysis software, such as OpenCV~\cite{OpenCV}. It contains useful functionalities to {\em e.g.\/} determine contour lines enclosing a particular flake and measure the corresponding enclosed area, fit an approximate polygonal shape or find the minimum desired polygonal shape that encloses a given set of pixels in the image. Given that TMDs typically exhibit a triangular shape, we employ triangles as the chosen polygonal shape, leading to an estimation of the corner positions. By carrying out this process for each of the layers in a flake, it is possible to estimate the twist angles between each pair of layers from their corner positions by simple trigonometric calculation.

Nevertheless, the use of image analysis software as described above has several disadvantages. First, it is not easily automated, requiring human intervention and thus resulting in a slow and tedious process that cannot be applied efficiently on a large scale. A better alternative is to train a DL model to directly predict the twist angles from the appropriately cropped experimental image of a twisted bilayer [see Figure~\ref{fig_1}(d-e)]. Training such a model would require a large database of preprocessed experimental images for which the twist angles had been previously determined. However there is a more practical and expedient procedure, which consists of employing a synthetically generated image database. As can be seen in Figure~\ref{fig_1} and Figure~\ref{fig_2}, experimental samples of $\mbox{MoS}_2$ bilayer flakes typically consist of two roughly triangular shapes, with the rotation angles in the range of 0 to $60^\circ$. It is trivial to generate large numbers of synthetic images of rotated pairs of superimposed triangles. This adds the advantage that such images can be made to sample homogeneously all possible twist angles, while training on labelled experimental images would result in a bias towards the experimentally observed twist angles. It is even possible to generate synthetic images that mimic more faithfully the experimental ones, by e.g. cropping corners of the triangles, or introducing random noise to their edges.

\begin{figure}[H]
\centering
\includegraphics[width=\textwidth]{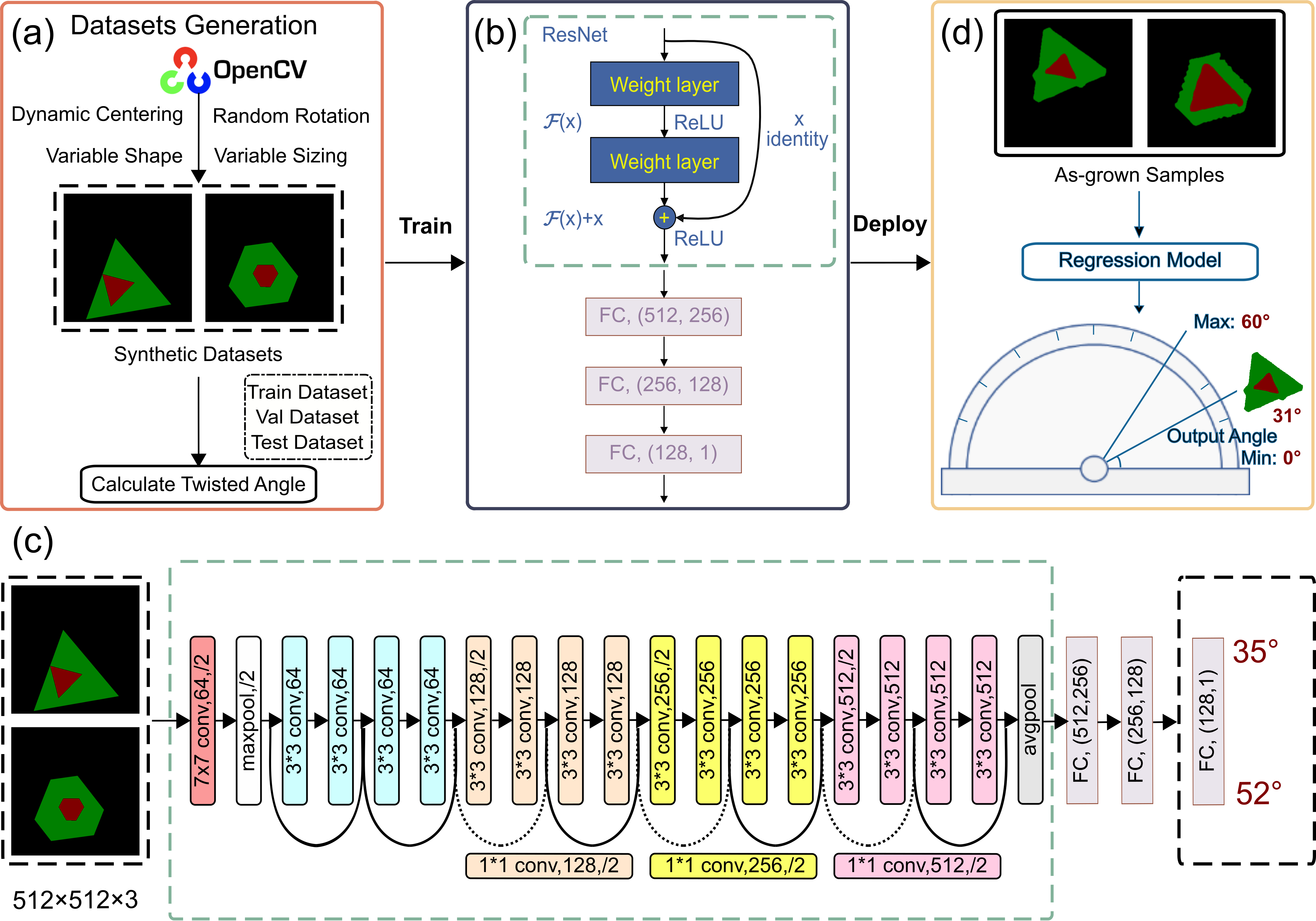}
\captionsetup{width=\textwidth}
\caption{\label{fig_3}
Deep learning approach for recognizing twist angles in atomically thin bilayer flakes. (a) Synthetic datasets illustrating varying twist angles in uniformly colored $\mbox{MoS}_2$ flakes post-segmentation. (b) ResNet CNN model training using the linear regression approach on the datasets from (a). (c) The detailed structure of ResNet with input training synthetic datasets and output twist angles of the datasets. (d) Prediction of twist angles for actual as-grown $\mbox{MoS}_2$ bilayer samples using the trained CNN model.}
\end{figure}

In the process of generating the datasets for training the second neural network, a sequence of images is created, including double-layered polygons that morph sequentially from hexagons to truncated triangles and finally into triangles (as illustrated in Figure~\ref{fig_3}(a)). Formation of the outer polygon begins with the careful selection of random variables, such as central coordinates, side lengths, and rotational angles, promoting visual diversity. Once the outer polygon is defined, an inner polygon is positioned within it. The inner polygon's location, rotation, and side length are determined through random selection, while ensuring its side length is equal to or less than the outer polygon's, preserving their hierarchical nesting and maintaining distinct spatial separation between them.

The angle formed by a vertex of each polygon and its respective center is calculated, providing a measure to assess the angular difference between them. This angular variation is then incorporated into the filename of the stored image, serving as a perceptible geometric attribute label. Consequently, the generated datasets, enriched with clear geometric annotations and offering a wide variety of forms, become a valuable resource for training datasets in the subsequent recognition of twist angles in CVD-grown samples.

\begin{figure}[H]
\centering
\includegraphics[width=\textwidth]{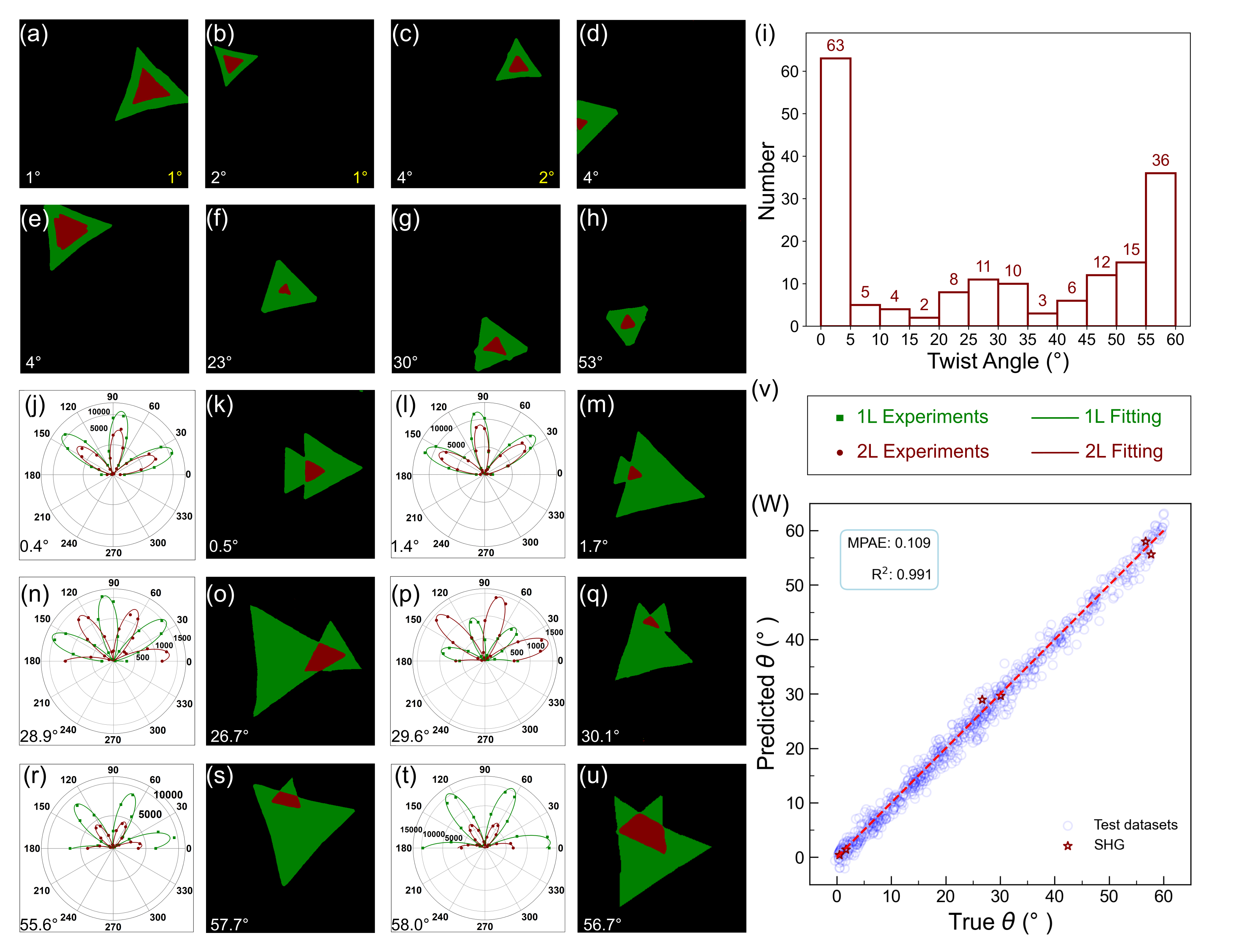}
\captionsetup{width=\textwidth}
\caption{\label{fig_4}
Performance evaluation of the twisted bilayer $\mbox{MoS}_2$ identification CNN model. (a)-(h) present predicted angles for different flakes, showcasing the efficacy of the CNN-based model (indicated in the bottom-left corner of each subfigure) in comparison to angles obtained via OpenCV (displayed in the bottom-right corner). It is noteworthy that in numerous instances, OpenCV was unable to identify the angles, resulting in absent data in the bottom-right corner of some subfigures. (i) Histogram of the quantity of bilayer $\mbox{MoS}_2$ at various twist angles. (j)-(u) Second harmonic generation (SHG) and thickness identification for corresponding samples with (v) serving as the legend. (w) Comparison between true and predicted twist angle angles $\theta$ using the artificially generated test datasets and SHG dataset in (j)-(u), respectively. }
\end{figure}

After that, a {\em Residual Network\/} (ResNet)~\cite{ResNet} Convolutional Neural Network (see Figure~\ref{fig_3}) is trained to predict the twist angle by regression to a synthetically generated database of more than 10000 images. The input images are RGB images with a resolution of 512 $\times$ 512 pixels obtained from Figure~\ref{fig_3}(a). In model design, the choice between using a single-layer or multi-layer approach for regression tasks hinges on the complexity of the data relationship. A direct 512 to 1 dimension reduction in the network might be suitable for simpler, linear problem, while a more gradual decrease, such as 512-256-128-1, tends to perform better for capturing complex, non-linear relationships in the data (as shown in Figure~\ref{fig_3} (b) and (c)). In the latter approach, additional layers aid the model in learning nuanced data patterns, offering a potential boost in predictive accuracy for intricate problems. Upon being input into the ResNet network, the final angles are obtained, as illustrated in Figure~\ref{fig_3} (b) and (c). After training the CNN regression model, the real as-grown bilayer $\mbox{MoS}_2$ micrographs after the first deep learning model are used as input, and the twist angle of the bilayer $\mbox{MoS}_2$ are obtained through this second neural network. The capabilities of the model are illustrated in Figure~\ref{fig_4}. The identification of twist bilayer $\mbox{MoS}_2$ using OpenCV is also demonstrated in the supplementary (Figure S8) and shown in the Figure~\ref{fig_4} as a comparison for the DL methodology. It can be concluded that the CNN regression model we used here can identify the twist angle in bilayer $\mbox{MoS}_2$ while being tolerant to shape irregularities.
For bilayer graphene synthesized through CVD, it is evident that different edges can correspond to varying twist angles in the secondary layer \cite{liu2021graphene} (Figure S15). Demonstrably, our methodology proficiently enables the identification of such twist angles within hexagonally shaped, CVD-grown bilayer graphene, offering insightful exploration into its structural intricacies, as depicted in Supplementary Figure S15.

To verify the identification of the twist angles of CVD-grown bilayer $\mbox{MoS}_2$, SHG measurements are performed on these samples with various twist angles as shown in Figure ~\ref{fig_4} (j)-(v).   The true and predicted twist angle $\theta$ using the artificially generated test datasets and SHG dataset is summarized in Figure~\ref{fig_4} (w), showing the reasonable accuracy of our model.

\begin{figure}[H]
\centering
\includegraphics[width=\textwidth]{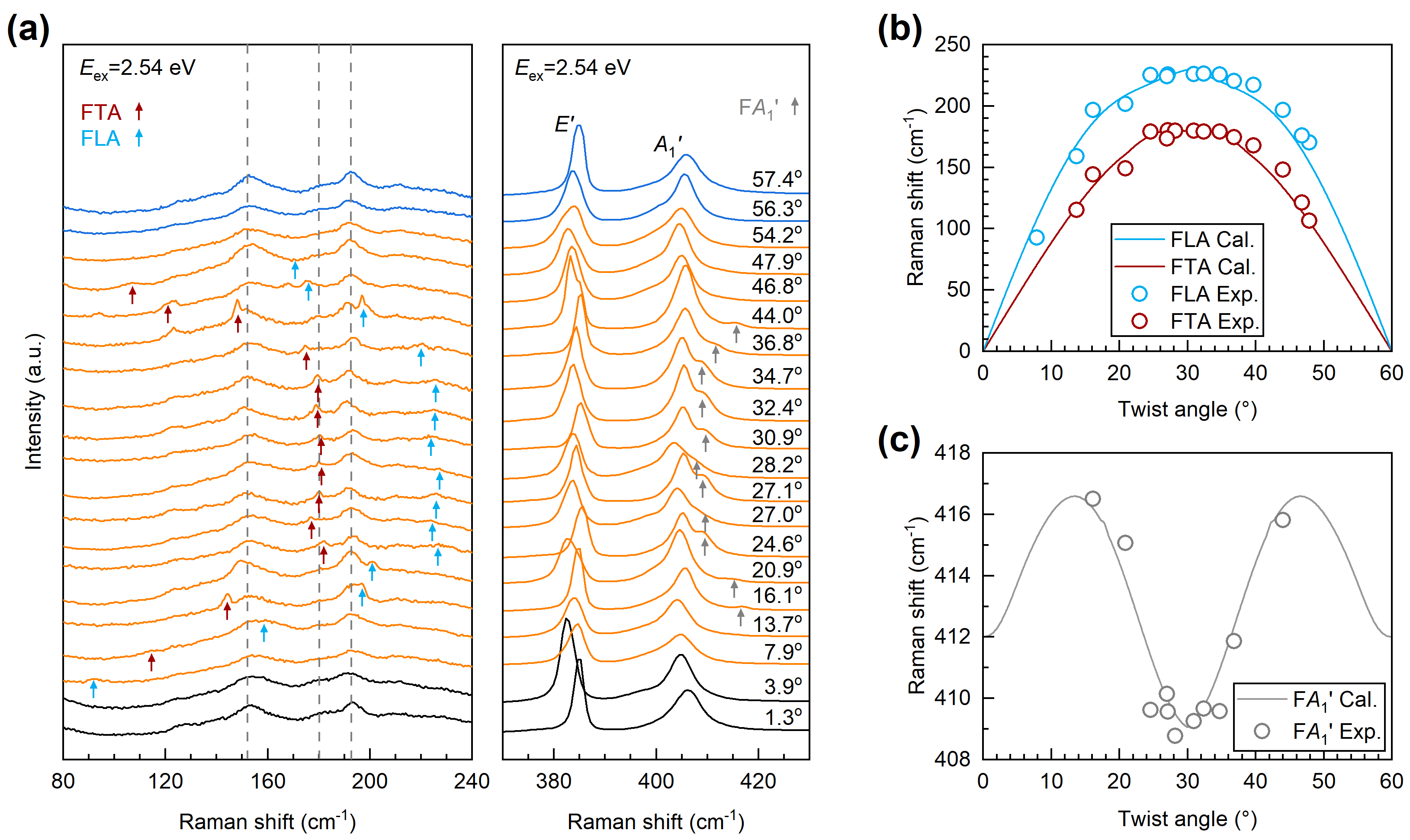}
\captionsetup{width=\textwidth}
\caption{\label{fig_5}
{Moir\'{e} phonons in twisted CVD-grown bilayer $\mbox{MoS}_2$. (a) Raman spectra of CVD grown bilayer $\mbox{MoS}_2$ with the excitation energy ($E_{\text{ex}}$) of 2.54 eV. Peaks assigned to moir\'{e} phonons modes~\cite{Lin2018Twisted} including FTA, FLA and F$A_{1}^{\prime}$ modes are marked by red, blue and gray arrows, respectively. The dashed lines corresponds to the peak position of second-order Raman modes and the $E^{\prime}$ and $A_{1}^{\prime}$ modes are also marked out. (b, c) Comparison between the calculated (Cal, lines) and experimental (Exp, circles) frequencies of twist-angle-dependent moir\'{e} phonons including (b) FTA, FLA and (c) F$A_{1}^{\prime}$ phonons.}}
\end{figure}

One possible application of our methodology could be used to select more samples for probing their optical properties correlations.  The strain, defects and doping of the inhomogeneity of the individual bilayer CVD-grown flakes could influence the Raman spectra, making the interpretation of Raman signal challenging\cite{lau2022reproducibility,lin2021large,lu2022unraveling,yang2022raman,xie2021straining,hu2022straining,sun2023abnormal}.
Figure ~\ref{fig_5} (a) shows the Raman spectra of the CVD-grown bilayer $\mbox{MoS}_2$ samples with twist angles ranging from 0-60$^{\circ}$. The three vertical dashed lines correspond to the peak positions of the second-order Raman modes $E^{\prime}$(M)$^{\text{LO}_{2}}$-LA(M) ($\sim$ 150 cm$^{-1}$), $A_{1}^{\prime}$(M)-LA(M) ($\sim$ 178 cm$^{-1}$) and TA(K) ($\sim$ 190 cm$^{-1}$)\cite{zhang2015phonon}. In addition, several branches of moir\'{e} phonons are observed in the Raman spectra. Moir\'{e} phonons refer to zone-center phonons in twisted bilayer $\mbox{MoS}_{2}$, which are folded from the off-center phonons in monolayer $\mbox{MoS}_{2}$ due to the periodic moir\'{e} potential and thus exhibit shift in peak frequency with the change of twist angle\cite{Lin2018Twisted}. The arrows indicates the assigned moir\'{e} phonon modes including folded TA (FTA), folded LA (FLA) and folded $A_{1}^{\prime}$ (F$A_{1}^{\prime}$) modes, which exhibit high frequency sensitivity to the twist angle, similar to the previous results\cite{Lin2018Twisted}. Figure~\ref{fig_5} (b)-(c) summarized the experimental and calculated frequencies of the three moir\'{e} phonon modes for comparison. The experimental peak positions of moir\'{e} phonons agree well with the theoretical ones\cite{Lin2018Twisted} based on twist angle of twisted bilayer $\mbox{MoS}_{2}$ and the phonon dispersion of monolayer $\mbox{MoS}_2$. Note that the frequencies of FA\textsubscript{1}$'$ mode on the CVD-grown bilayer MoS\textsubscript{2} diverging relatively from the theoretical curve in the range of 24.6\textdegree{} to 34.7\textdegree{} may be arisen from different reasons, where strain is very likely as E$'$ and A\textsubscript{1}$'$ modes also show behaviors of strain-induced shifts \cite{yang2022raman}. We did not discuss the results of twist angle at 0-5$^{\circ}$ and 55-60$^{\circ}$ in detail, because the case is very complicated due to the phonon renormalization induced by lattice relaxation\cite{quan2021phonon}.

Another possible application of our methodology could be used for optimization of the large-scale growth for bilayer atomically materials (as shown in Figure S17).  In addition, the exfoliated flakes with regular shape could also be the next possible applications of our model (as show in Figure S18) \cite{Palai2023,zhang2018moire}.  This thorough work-flow acts as a clear guide, detailing the methodology utilized in the study for the precise identification and analysis of twisted bilayers in TMDs grown via CVD. This method can also be adapted to include graphene (as shown in Figure S15) and hBN, as well as their heterostructures.

In conclusion, we have demonstrated a robust and efficient methodology for the identification and analysis of twisted bilayers in TMDs using deep learning and OpenCV techniques. Our approach, which combines optical micrographs, deep learning models, and OpenCV, provides accurate and comprehensive predictions of thickness properties and twist angles of bilayer TMDs. The comparison of various deep learning models revealed that the {\em U-Net\/} model exhibits superior performance in terms of global accuracy, mean intersection over union, and processing speed.

Our methodology can be extended to other two-dimensional materials grown by CVD, including both homojunctions and heterojunctions, highlighting its versatility and broad applicability in the field of 2D materials analysis. Our datasets and codes are made freely available as a service to the community. We hope that, by facilitating and automating the structural analysis of TMDs, this work will contribute to further advancements in the field of TMDs and 2D materials in general, thus also the rapid growing Autonomous lab using AI\cite{ren2023autonomous}.

\section{Methods}
\label{sec:methods}
A description of the data pipeline, data preparation, deep learning training of the semantic segmentation model, twist angle identification by OpenCV can be found in SI.

\section{Author Contributions}

\mbox{Y. X.} supervised the project; \mbox{Y. X.} and \mbox{E. R. H.} conceived the idea. \mbox{H. Y.}, \mbox{R. H.}, \mbox{H. W.}, and \mbox{X. H.} contributed equally to this work; \mbox{H. Y.} performed the coding with the help of \mbox{Y. X.} and \mbox{R. H.}; \mbox{H. Y.} and \mbox{Y. Z. X.} prepared the datasets with assistance from \mbox{H. C.} and \mbox{M. G.}; \mbox{H. Y.}, \mbox{X. H.}, \mbox{K. H.}, and \mbox{Y. Z.} prepared the figures with the help of \mbox{Y. X.}, \mbox{R. H.}, and \mbox{E. R. H.}; \mbox{H. W.}, \mbox{Y. Z.}, and \mbox{X. Z.} measured the Raman spectra and SHG, and analyzed the data under the supervision of \mbox{P. T.}; \mbox{Y. X.}, \mbox{R. H.}, and \mbox{E. R. H.} wrote the manuscript.

\section{Acknowledgements}
 This work received funding from the National Natural Science Foundation of China (NSFC) Grants (No. 62011530438, No. 61704129, No.12204472, No. 12127807) are acknowledged. This work was partially supported by the Key Research and Development Program of Shaanxi (Program No.2021KW-02), Fundamental Research Funds for the Central Universities (QTZX23026), and the fund of the State Key Laboratory of Solidification Processing in Northwestern Polytechnical University (grant no. SKLSP201612). The work of ERH is supported by MCIN/AEI/ 10.13039/501100011033/FEDER, UE through projects PID2022-139776NB-C66 and TED2021-132267B-I00.

\section{Supporting Information}

This material is available free of charge via the Internet at \href{http://pubs.acs.org}{ACS Publications}. It includes detailed descriptions of dataset preparation, semantic segmentation results, synthetic dataset generation, verification of twisted bilayer graphene, and Raman spectra of CVD-grown twisted bilayer $\mathrm{MoS_2}$.

All scripts used in this study are available on GitHub~\cite{Xie2024Twist2DNet}. Currently, the source code is provided as a supplementary file for reviewers and readers to examine. Upon acceptance of the manuscript, the code will be fully open-sourced at the following repository: \url{https://github.com/YongXie-ICMM/Twist2DNet}.

Each subfolder in the supplementary package corresponds directly to the respective sections described in this Supporting Information.

\bibliography{./Bibliography.bib}

@article{Cao2018magicangle,
   author = {Cao, Yuan and Fatemi, Valla and Fang, Shiang and Watanabe, Kenji and Taniguchi, Takashi and Kaxiras, Efthimios and Jarillo-Herrero, Pablo},
   title = {Unconventional superconductivity in magic-angle graphene superlattices},
   journal = {Nature},
   volume = {556},
   number = {7699},
   pages = {43-50},
   ISSN = {1476-4687},
   DOI = {10.1038/nature26160},
   url = {https://doi.org/10.1038/nature26160},
   year = {2018},
   type = {Journal Article}
}

@article{Cao2018correlated,
   author = {Cao, Yuan and Fatemi, Valla and Demir, Ahmet and Fang, Shiang and Tomarken, Spencer L. and Luo, Jason Y. and Sanchez-Yamagishi, Javier D. and Watanabe, Kenji and Taniguchi, Takashi and Kaxiras, Efthimios and Ashoori, Ray C. and Jarillo-Herrero, Pablo},
   title = {Correlated insulator behaviour at half-filling in magic-angle graphene superlattices},
   journal = {Nature},
   volume = {556},
   number = {7699},
   pages = {80-84},
   ISSN = {1476-4687},
   DOI = {10.1038/nature26154},
   url = {https://doi.org/10.1038/nature26154},
   year = {2018},
   type = {Journal Article}
}

@article{tang2020simulation,
  title={Simulation of Hubbard model physics in $\mbox{WSe}_2$/$\mbox{WS}_2$ moir{\'e} superlattices},
  author={Tang, Yanhao and Li, Lizhong and Li, Tingxin and Xu, Yang and Liu, Song and Barmak, Katayun and Watanabe, Kenji and Taniguchi, Takashi and MacDonald, Allan H and Shan, Jie and Mak, Kin Fai},
  journal={Nature},
  volume={579},
  number={7799},
  pages={353--358},
  year={2020},
  publisher={Nature Publishing Group UK London}
}

@article{wu2018hubbard,
  title={Hubbard model physics in transition metal dichalcogenide moir{\'e} bands},
  author={Wu, Fengcheng and Lovorn, Timothy and Tutuc, Emanuel and MacDonald, Allan H},
  journal={Phys. Rev. Lett.},
  volume={121},
  number={2},
  pages={026402},
  year={2018},
  publisher={APS}
}

@article{wang2020correlated,
  title={Correlated electronic phases in twisted bilayer transition metal dichalcogenides},
  author={Wang, Lei and Shih, En-Min and Ghiotto, Augusto and Xian, Lede and Rhodes, Daniel A and Tan, Cheng and Claassen, Martin and Kennes, Dante M and Bai, Yusong and Kim, Bumho and Watanabe, Kenji and Taniguchi, Takashi and Zhu, Xiaoyang and Hone, James and  and Rubio, Angel and  Pasupathy, Abhay N. and Dean, Cory R.},
  journal={Nat. Mater.},
  volume={19},
  number={8},
  pages={861--866},
  year={2020},
  publisher={Nature Publishing Group UK London}
}

@article{devakul2021magic,
  title={Magic in twisted transition metal dichalcogenide bilayers},
  author={Devakul, Trithep and Cr{\'e}pel, Valentin and Zhang, Yang and Fu, Liang},
  journal={Nat. Commun.},
  volume={12},
  number={1},
  pages={6730},
  year={2021},
  publisher={Nature Publishing Group UK London}
}

@article{Lau2022review,
   author = {Lau, Chun Ning and Bockrath, Marc W. and Mak, Kin Fai and Zhang, Fan},
   title = {Reproducibility in the fabrication and physics of moir\'{e} materials},
   journal = {Nature},
   volume = {602},
   number = {7895},
   pages = {41-50},
   ISSN = {1476-4687},
   DOI = {10.1038/s41586-021-04173-z},
   url = {https://doi.org/10.1038/s41586-021-04173-z},
   year = {2022},
   type = {Journal Article}
}

@article{quan2021phonon,
  title={Phonon renormalization in reconstructed $\mbox{MoS}_2$ moir{\'e} superlattices},
  author={Quan, Jiamin and Linhart, Lukas and Lin, Miao-Ling and Lee, Daehun and Zhu, Jihang and Wang, Chun-Yuan and Hsu, Wei-Ting and Choi, Junho and Embley, Jacob and Young, Carter and Taniguchi, Takashi and Watanabe, Kenji and Shih, Chih-Kang and Lai, Keji and MacDonald, Allan H. and Tan, Ping-Heng and  Libisch, Florian and Li, Xiaoqin
},
  journal={Nat. Mater.},
  volume={20},
  number={8},
  pages={1100--1105},
  year={2021},
  publisher={Nature Publishing Group UK London}
}

@article{marcellina2021evidence,
  title={Evidence for moir{\'e} trions in twisted $\mbox{MoSe}_2$ homobilayers},
  author={Marcellina, Elizabeth and Liu, Xue and Hu, Zehua and Fieramosca, Antonio and Huang, Yuqing and Du, Wei and Liu, Sheng and Zhao, Jiaxin and Watanabe, Kenji and Taniguchi, Takashi and Xiong, Qihua},
  journal={Nano Lett.},
  volume={21},
  number={10},
  pages={4461--4468},
  year={2021},
  publisher={ACS Publications}
}

@article{lin2021twist,
  title={Twist-angle engineering of excitonic quantum interference and optical nonlinearities in stacked 2D semiconductors},
  author={Lin, Kai-Qiang and Faria Junior, Paulo E and Bauer, Jonas M and Peng, Bo and Monserrat, Bartomeu and Gmitra, Martin and Fabian, Jaroslav and Bange, Sebastian and Lupton, John M},
  journal={Nat. Commun.},
  volume={12},
  number={1},
  pages={1553},
  year={2021},
  publisher={Nature Publishing Group UK London}
}

@article{peimyoo2021electrical,
  title={Electrical tuning of optically active interlayer excitons in bilayer $\mbox{MoS}_2$},
  author={Peimyoo, Namphung and Deilmann, Thorsten and Withers, Freddie and Escolar, Janire and Nutting, Darren and Taniguchi, Takashi and Watanabe, Kenji and Taghizadeh, Alireza and Craciun, Monica Felicia and Thygesen, Kristian Sommer and Russo, Saverio},
  journal={Nat. Nanotechnol.},
  volume={16},
  number={8},
  pages={888--893},
  year={2021},
  publisher={Nature Publishing Group UK London}
}

@article{castellanos2014deterministic,
  title={Deterministic transfer of two-dimensional materials by all-dry viscoelastic stamping},
  author={Castellanos-Gomez, Andres and Buscema, Michele and Molenaar, Rianda and Singh, Vibhor and Janssen, Laurens and Van Der Zant, Herre SJ and Steele, Gary A},
  journal={2D Mater.},
  volume={1},
  number={1},
  pages={011002},
  year={2014},
  publisher={IOP Publishing}
}

@article{wang2013one,
  title={One-dimensional electrical contact to a two-dimensional material},
  author={Wang, Lei and Meric, I and Huang, PY and Gao, Q and Gao, Y and Tran, H and Taniguchi, T and Watanabe, Kenji and Campos, LM and Muller, DA and Guo, J and Kim, P and Hone, J and Shepard, KL and Dean, CR},
  journal={Science},
  volume={342},
  number={6158},
  pages={614--617},
  year={2013},
  publisher={American Association for the Advancement of Science}
}

@article{cheng2015kinetic,
  title={Kinetic nature of grain boundary formation in as-grown $\mbox{MoS}_2$ monolayers},
  author={Cheng, Jingxin and Jiang, Tao and Ji, Qingqing and Zhang, Yu and Li, Zhiming and Shan, Yuwei and Zhang, Yanfeng and Gong, Xingao and Liu, Weitao and Wu, Shiwei},
  journal={Adv. Mater.},
  volume={27},
  number={27},
  pages={4069--4074},
  year={2015},
  publisher={Wiley Online Library}
}

@article{Xie2017,
   author = {Xie, Yong and Wang, Zhan and Zhan, Yongjie and Zhang, Peng and Wu, Ruixue and Jiang, Teng and Wu, Shiwei and Wang, Hong and Zhao, Ying and Nan, Tang and Ma, Xiaohua},
   title = {Controllable growth of monolayer $\mbox{MoS}_2$ by chemical vapor deposition via close $\mbox{MoO}_2$ precursor for electrical and optical applications},
   journal = {Nanotechnology},
   volume = {28},
   number = {8},
   pages = {084001},
   ISSN = {1361-6528},
   url = {http://iopscience.iop.org/10.1088/1361-6528/aa5439},
   year = {2017},
   type = {Journal Article}
}

@article{paradisanos2020,
  title={Controlling interlayer excitons in $\mbox{MoS}_2$ layers grown by chemical vapor deposition},
  author={Paradisanos, Ioannis and Shree, Shivangi and George, Antony and Leisgang, Nadine and Robert, Cedric and Watanabe, Kenji and Taniguchi, Takashi and Warburton, Richard J and Turchanin, Andrey and Marie, Xavier and Gerber, Iann C and Urbaszek, Bernhard},
  journal={Nat. Commun.},
  volume={11},
  number={1},
  pages={2391},
  year={2020},
  publisher={Nature Publishing Group UK London}
}

@article{Lin2018Twisted,
   author = {Lin, Miao-Ling and Tan, Qing-Hai and Wu, Jiang-Bin and Chen, Xiao-Shuang and Wang, Jin-Huan and Pan, Yu-Hao and Zhang, Xin and Cong, Xin and Zhang, Jun and Ji, Wei and Hu, Ping-An and Liu, Kai-Hui and Tan, Ping-Heng},
   title = {Moiré Phonons in Twisted Bilayer $\mbox{MoS}_2$},
   journal = {ACS Nano},
   volume = {12},
   number = {8},
   pages = {8770-8780},
   ISSN = {1936-0851},
   DOI = {10.1021/acsnano.8b05006},
   url = {https://doi.org/10.1021/acsnano.8b05006},
   year = {2018},
   type = {Journal Article}
}

@article{wu2023analyzing,
author = {Wu, Heng and Lin, Miao-Ling and Zhou, Yan and Zhang, Xin and Tan, Ping-Heng},
   title = {Analyzing Fundamental Properties of Two-Dimensional Materials by Raman Spectroscopy from Microscale to Nanoscale},
   journal = {Analytical Chemistry},
   volume = {95},
   number = {29},
   pages = {10821-10838},
   ISSN = {0003-2700},
   DOI = {10.1021/acs.analchem.3c00272},
   url = {https://doi.org/10.1021/acs.analchem.3c00272},
   year = {2023},
   type = {Journal Article},
  publisher={ACS Publications}
}

@article{zhang2019transition,
 title={Transition metal dichalcogenides bilayer single crystals by reverse-flow chemical vapor epitaxy},
  author={Zhang, Xiumei and Nan, Haiyan and Xiao, Shaoqing and Wan, Xi and Gu, Xiaofeng and Du, Aijun and Ni, Zhenhua and Ostrikov, Kostya},
  journal={Nat. Commun.},
  volume={10},
  number={1},
  pages={598},
  year={2019},
  publisher={Nature Publishing Group UK London}
}

@article{Yin2014,
   author = {Yin, Xiaobo and Ye, Ziliang and Chenet, Daniel A. and Ye, Yu and O{'}Brien, Kevin and Hone, James C. and Zhang, Xiang},
   title = {Edge Nonlinear Optics on a $\mbox{MoS}_2$ Atomic Monolayer},
   journal = {Science},
   volume = {344},
   number = {6183},
   pages = {488-490},
   doi = {10.1126/science.1250564},
   url = {https://www.science.org/doi/abs/10.1126/science.1250564},
   year = {2014}
}

@article{PhysRevB.87.161403,
  title = {Second harmonic microscopy of monolayer MoS${}_{2}$},
  author = {Kumar, Nardeep and Najmaei, Sina and Cui, Qiannan and Ceballos, Frank and Ajayan, Pulickel M. and Lou, Jun and Zhao, Hui},
  journal = {Phys. Rev. B},
  volume = {87},
  issue = {16},
  pages = {161403},
  numpages = {6},
  year = {2013},
  month = {Apr},
  publisher = {American Physical Society},
  doi = {10.1103/PhysRevB.87.161403},
  url = {https://link.aps.org/doi/10.1103/PhysRevB.87.161403}
}

@article{Carrascoso2020,
doi = {10.1088/2515-7639/ab4432},
url = {https://dx.doi.org/10.1088/2515-7639/ab4432},
year = {2019},
month = {oct},
publisher = {IOP Publishing},
volume = {3},
number = {1},
pages = {015003},
author = {Felix Carrascoso and Der-Yuh Lin and Riccardo Frisenda and Andres Castellanos-Gomez},
title = {Biaxial strain tuning of interlayer excitons in bilayer $\mbox{MoS}_2$},
journal = {J. Phys. Materials},

}

@article{Zande2014Twist,
   author = {van der Zande, Arend M. and Kunstmann, Jens and Chernikov, Alexey and Chenet, Daniel A. and You, YuMeng and Zhang, XiaoXiao and Huang, Pinshane Y. and Berkelbach, Timothy C. and Wang, Lei and Zhang, Fan and Hybertsen, Mark S. and Muller, David A. and Reichman, David R. and Heinz, Tony F. and Hone, James C.},
   title = {Tailoring the Electronic Structure in Bilayer Molybdenum Disulfide via Interlayer Twist},
   journal = {Nano Lett.},
   volume = {14},
   number = {7},
   pages = {3869-3875},
   ISSN = {1530-6984},
   DOI = {10.1021/nl501077m},
   url = {https://doi.org/10.1021/nl501077m},
   year = {2014},
   type = {Journal Article}
}

@article{Tan-2012-NM,
	doi = {10.1038/NMAT3245},
	year = {2012},
	volume = {11},
	pages = {294-300},
	author = {Tan, P. H. and Han, W. P. and Zhao, W. J. and Wu, Z. H. and Chang, K. and Wang, H. and Wang, Y. F. and Bonini, N. and Marzari, N. and Pugno, N. and Savini, G. and Lombardo, A. and Ferrari, A. C.},
	title = {The shear mode of multilayer graphene},
	journal = {Nature Materials}
}

@article{Zhang-2013-PRB,
	doi = {10.1103/PhysRevB.87.115413},
	year = {2013},
	volume = {87},
	pages = {115413},
	author = {Zhang, X. and Han, W. P. and Wu, J. B. and Milana, S. and Lu, Y. and Li, Q. Q. and Ferrari, A. C. and Tan, P. H.},
	title = {Raman spectroscopy of shear and layer breathing modes in multilayer {M}o{S}$_2$},
	journal = {Phys. Rev. B}
}

@article{Yoo2019bilayergraphene,
   author = {Yoo, Hyobin and Engelke, Rebecca and Carr, Stephen and Fang, Shiang and Zhang, Kuan and Cazeaux, Paul and Sung, Suk Hyun and Hovden, Robert and Tsen, Adam W. and Taniguchi, Takashi and Watanabe, Kenji and Yi, Gyu-Chul and Kim, Miyoung and Luskin, Mitchell and et al.},
   title = {Atomic and electronic reconstruction at the van der Waals interface in twisted bilayer graphene},
   journal = {Nat. Mater.},
   volume = {18},
   number = {5},
   pages = {448-453},
   ISSN = {1476-4660},
   DOI = {10.1038/s41563-019-0346-z},
   url = {https://doi.org/10.1038/s41563-019-0346-z},
   year = {2019},
   type = {Journal Article}
}

@article{Liu2015Bilayer,
   author = {Liu, Hongjun and Zheng, Hao and Yang, Fang and Jiao, Lu and Chen, Jinglei and Ho, Wingkin and Gao, Chunlei and Jia, Jinfeng and Xie, Maohai},
   title = {Line and point defects in $\mbox{MoSe}_2$ bilayer studied by scanning tunneling microscopy and spectroscopy},
   journal = {ACS Nano},
   volume = {9},
   number = {6},
   pages = {6619-6625},
   ISSN = {1936-0851},
   DOI = {10.1021/acsnano.5b02789},
   url = {https://doi.org/10.1021/acsnano.5b02789},
   year = {2015},
   type = {Journal Article}
}

@article{li2012bulk,
  title={From bulk to monolayer $\mbox{MoS}_2$: evolution of Raman scattering},
  author={Li, Hong and Zhang, Qing and Yap, Chin Chong Ray and Tay, Beng Kang and Edwin, Teo Hang Tong and Olivier, Aurelien and Baillargeat, Dominique},
  journal={Advanced Functional Materials},
  volume={22},
  number={7},
  pages={1385--1390},
  year={2012},
  publisher={Wiley Online Library}
}

@article{castellanos2010optical,
  title={Optical identification of atomically thin dichalcogenide crystals},
  author={Castellanos-Gomez, Andres and Agra{\"\i}t, Nicolas and Rubio-Bollinger, Gabino},
  journal={Applied Physics Letters},
  volume={96},
  number={21},
  pages={213116},
  year={2010},
  publisher={AIP Publishing}
}

@article{dumcenco2015large,
  title={Large-area epitaxial monolayer $\mbox{MoS}_2$},
  author={Dumcenco, Dumitru and Ovchinnikov, Dmitry and Marinov, Kolyo and Lazic, Predrag and Gibertini, Marco and Marzari, Nicola and Sanchez, Oriol Lopez and Kung, Yen-Cheng and Krasnozhon, Daria and Chen, Ming-Wei and Bertolazzi, Simone and Gillet, Philippe and Morral, Anna Fontcuberta i and Radenovic, Aleksandra and Kis, Andras},
  journal={ACS Nano},
  volume={9},
  number={4},
  pages={4611--4620},
  year={2015},
  publisher={ACS Publications}
}

@article{frisenda2018robotic,
  title={Robotic assembly of artificial nanomaterials},
  author={Frisenda, Riccardo and Castellanos-Gomez, Andres},
  journal={Nat. Nanotechnol.},
  volume={13},
  number={6},
  pages={441--442},
  year={2018},
  publisher={Nature Publishing Group UK London}
}

@article{masubuchi2018autonomous,
  title={Autonomous robotic searching and assembly of two-dimensional crystals to build van der Waals superlattices},
  author={Masubuchi, Satoru and Morimoto, Masataka and Morikawa, Sei and Onodera, Momoko and Asakawa, Yuta and Watanabe, Kenji and Taniguchi, Takashi and Machida, Tomoki},
  journal={Nat. Commun.},
  volume={9},
  number={1},
  pages={1413},
  year={2018},
  publisher={Nature Publishing Group UK London}
}

@article{lin2018intelligent,
  title={Intelligent identification of two-dimensional nanostructures by machine-learning optical microscopy},
  author={Lin, Xiaoyang and Si, Zhizhong and Fu, Wenzhi and Yang, Jianlei and Guo, Side and Cao, Yuan and Zhang, Jin and Wang, Xinhe and Liu, Peng and Jiang, Kaili and Zhao, Weisheng},
  journal={Nano Res.},
  volume={11},
  pages={6316--6324},
  year={2018},
  publisher={Springer}
}

@article{masubuchi2019classifying,
  title={Classifying optical microscope images of exfoliated graphene flakes by data-driven machine learning},
  author={Masubuchi, Satoru and Machida, Tomoki},
  journal={npj 2D Mater. Appl.},
  volume={3},
  number={1},
  pages={4},
  year={2019},
  publisher={Nature Publishing Group UK London}
}

@article{han2020deep,
  title={Deep-learning-enabled fast optical identification and characterization of 2D materials},
  author={Han, Bingnan and Lin, Yuxuan and Yang, Yafang and Mao, Nannan and Li, Wenyue and Wang, Haozhe and Yasuda, Kenji and Wang, Xirui and Fatemi, Valla and Zhou, Lin and Wang, Joel I.-Jan and Ma, Qiong and Cao, Yuan and Rodan-Legrain, Daniel and et al.},
  journal={Adv. Mater.},
  volume={32},
  number={29},
  pages={2000953},
  year={2020},
  publisher={Wiley Online Library}
}

@article{masubuchi2020deep,
  title={Deep-learning-based image segmentation integrated with optical microscopy for automatically searching for two-dimensional materials},
  author={Masubuchi, Satoru and Watanabe, Eisuke and Seo, Yuta and Okazaki, Shota and Sasagawa, Takao and Watanabe, Kenji and Taniguchi, Takashi and Machida, Tomoki},
  journal={npj 2D Mater. Appl.},
  volume={4},
  number={1},
  pages={3},
  year={2020},
  publisher={Nature Publishing Group UK London}
}

@article{sterbentz2021universal,
  title={Universal image segmentation for optical identification of 2D materials},
  author={Sterbentz, Randy M and Haley, Kristine L and Island, Joshua O},
  journal={Sci. Rep.},
  volume={11},
  number={1},
  pages={5808},
  year={2021},
  publisher={Nature Publishing Group UK London}
}

@article{liu2021graphene,
  title={Hetero-site nucleation for growing twisted bilayer graphene with a wide range of twist angles},
  author = {Sun, Luzhao and Wang, Zihao and Wang, Yuechen and Zhao, Liang and Li, Yanglizhi and Chen,Buhang and Huang, Shenghong and Zhang, Shishu and Wang,Wendong and Pei, Ding and Fang, Hongwei and Zhong, Shan and Liu, Haiyang and Zhang, Jincan and et al.},
  journal ={Nat. Commun.},
  volume={12},
  number={1},
  pages={2391},
  year = {2021},
  publisher={Nature Publishing Group UK London}
}

@misc{OpenCV,
  title = {Open Source Computer Vision Library},
  howpublished = {\url{https://opencv.org/}},
  note = {Accessed: February 10, 2024}
}

@article{wang2017nacl,
  title={NaCl-assisted one-step growth of $\mbox{MoS}_2$--$\mbox{WS}_2$ in-plane heterostructures},
  author={Wang, Zhan and Xie, Yong and Wang, Haolin and Wu, Ruixue and Nan, Tang and Zhan, Yongjie and Sun, Jing and Jiang, Teng and Zhao, Ying and Lei, Yimin and Yang, Mei and Wang, Weidong and Zhu, Qing and Ma, Xiaohua and Hao, Yue},
  journal={Nanotechnology},
  volume={28},
  number={32},
  pages={325602},
  year={2017},
  publisher={IOP Publishing}
}

@article{WANG2020144371,
title = {2H/1T$'$ phase WS$_2(1-x)$Te$_{2x}$ alloys grown by chemical vapor deposition with tunable band structures},
journal = {Applied Surface Science},
volume = {504},
pages = {144371},
year = {2020},
issn = {0169-4332},
doi = {https://doi.org/10.1016/j.apsusc.2019.144371},
url = {https://www.sciencedirect.com/science/article/pii/S0169433219331873},
author = {Zhan Wang and Jing Sun and Haolin Wang and Yimin Lei and Yong Xie and Guanfei Wang and Ying Zhao and Xiaobo Li and Hua Xu and Xiubo Yang and Liping Feng and Xiaohua Ma},
}

@misc{chen2017rethinking,
  title={Rethinking atrous convolution for semantic image segmentation},
  author={Chen, Liang-Chieh and Papandreou, George and Schroff, Florian and Adam, Hartwig},
  journal={arXiv preprint arXiv:1706.05587},
  year={2017},
  note = {accessed on December 06, 2023},
}

@inproceedings{long2015fully,
  title={Fully convolutional networks for semantic segmentation},
  author={Long, Jonathan and Shelhamer, Evan and Darrell, Trevor},
  booktitle={Proceedings of the IEEE conference on computer vision and pattern recognition},
  pages={3431--3440},
  year={2015}
}

@inproceedings{howard2019searching,
  title={Searching for mobilenetv3},
  author={Howard, Andrew and Sandler, Mark and Chu, Grace and Chen, Liang-Chieh and Chen, Bo and Tan, Mingxing and Wang, Weijun and Zhu, Yukun and Pang, Ruoming and Vasudevan, Vijay and Le, Quoc V. and Adam, Hartwig},
  booktitle={Proceedings of the IEEE/CVF international conference on computer vision},
  pages={1314--1324},
  year={2019}
}

@inproceedings{ronneberger2015u,
  title={U-net: Convolutional networks for biomedical image segmentation},
  author={Ronneberger, Olaf and Fischer, Philipp and Brox, Thomas},
  booktitle={Medical Image Computing and Computer-Assisted Intervention--MICCAI 2015: 18th International Conference, Munich, Germany, October 5-9, 2015, Proceedings, Part III 18},
  pages={234--241},
  year={2015},
  organization={Springer}
}

@inproceedings{ResNet,
  title={Deep residual learning for image recognition},
  author={He, Kaiming and Zhang, Xiangyu and Ren, Shaoqing and Sun, Jian},
  booktitle={Proceedings of the IEEE conference on computer vision and pattern recognition},
  pages={770--778},
  year={2016}
}

@software{torchvision2016,
  title        = {TorchVision: PyTorch's Computer Vision Library},
  author       = {{TorchVision maintainers and contributors}},
  year         = {2016},
  journal      = {GitHub repository},
  publisher    = {GitHub},
  howpublished = {\url{https://github.com/pytorch/vision}},
  note         = {Accessed: February 10, 2024}
}

@article{lau2022reproducibility,
  title={Reproducibility in the fabrication and physics of moir{\'e} materials},
  author={Lau, Chun Ning and Bockrath, Marc W and Mak, Kin Fai and Zhang, Fan},
  journal={Nature},
  volume={602},
  number={7895},
  pages={41--50},
  year={2022},
  publisher={Nature Publishing Group UK London}
}

@article{lin2021large,
  title={Large-scale mapping of moir{\'e} superlattices by hyperspectral Raman imaging},
  author={Lin, Kai-Qiang and Holler, Johannes and Bauer, Jonas M and Parzefall, Philipp and Scheuck, Marten and Peng, Bo and Korn, Tobias and Bange, Sebastian and Lupton, John M and Sch{\"u}ller, Christian},
  journal={Advanced Materials},
  volume={33},
  number={34},
  pages={2008333},
  year={2021},
  publisher={Wiley Online Library}
}

@article{lu2022unraveling,
  title={Unraveling the Correlation between Raman and Photoluminescence in Monolayer MoS$_2$ through Machine-Learning Models},
  author={Lu, Ang-Yu and Martins, Luiz Gustavo Pimenta and Shen, Pin-Chun and Chen, Zhantao and Park, Ji-Hoon and Xue, Mantian and Han, Jinchi and Mao, Nannan and Chiu, Ming-Hui and Palacios, Tom{\'a}s and Tung, Vincent and Kong, Jing
},
  journal={Advanced Materials},
  volume={34},
  number={34},
  pages={2202911},
  year={2022},
  publisher={Wiley Online Library}
}

@article{yang2022raman,
  title={Raman spectroscopic probe for nonlinear MoS$_2$ nanoelectromechanical resonators},
  author={Yang, Rui and Yousuf, SM Enamul Hoque and Lee, Jaesung and Zhang, Pengcheng and Liu, Zuheng and Feng, Philip X-L},
  journal={Nano Letters},
  volume={22},
  number={14},
  pages={5780--5787},
  year={2022},
  publisher={ACS Publications}
}

@article{xie2021straining,
  title={Straining and Tuning Atomic Layer Nanoelectromechanical Resonators via Comb-Drive MEMS Actuators},
  author={Xie, Yong and Lee, Jaesung and Wang, Yanan and Feng, Philip X-L},
  journal={Advanced Materials Technologies},
  volume={6},
  number={2},
  pages={2000794},
  year={2021},
  publisher={Wiley Online Library}
}

@article{hu2022straining,
  title={Straining of atomically thin  WSe$_2$ crystals: Suppressing slippage by thermal annealing},
  author={Hu, Wenshuai and Wang, Yabin and He, Kexin and He, Xiaolong and Bai, Yan and Liu, Chenyang and Zhou, Nan and Wang, Haolin and Li, Peixian and Ma, Xiaohua and Xie, Yong},
  journal={Journal of Applied Physics},
  volume={132},
  number={8},
 pages={085104},
  year={2022},
  publisher={AIP Publishing}
}

@article{zhang2015phonon,
	title = {Phonon and {Raman} scattering of two-dimensional transition metal dichalcogenides from monolayer, multilayer to bulk material},
	volume = {44},
	number = {9},
	journal = {Chem. Soc. Rev.},
	author = {Zhang, Xin and Qiao, Xiao-Fen and Shi, Wei and Wu, Jiang-Bin and Jiang, De-Sheng and Tan, Ping-Heng},
	year = {2015},
	pages = {2757--2785}
}

@article{Palai2023,
   author = {Palai, Swaroop Kumar and Dyksik, Mateusz and Sokolowski, Nikodem and Ciorga, Mariusz and Sánchez Viso, Estrella and Xie, Yong and Schubert, Alina and Taniguchi, Takashi and Watanabe, Kenji and Maude, Duncan K. and Surrente, Alessandro and Baranowski, Michał and Castellanos-Gomez, Andres and Munuera, Carmen and Plochocka, Paulina},
   title = {Approaching the Intrinsic Properties of Moiré Structures Using Atomic Force Microscopy Ironing},
   journal = {Nano Letters},
   volume = {23},
   number = {11},
   pages = {4749-4755},
   ISSN = {1530-6984},
   DOI = {10.1021/acs.nanolett.2c04765},
   url = {https://doi.org/10.1021/acs.nanolett.2c04765},
   year = {2023},
   type = {Journal Article}
}

@article{zhang2018moire,
  title={Moir{\'e} intralayer excitons in a MoSe$_2$/MoS$_2$ heterostructure},
  author={Zhang, Nan and Surrente, Alessandro and Baranowski, Micha{\l} and Maude, Duncan K and Gant, Patricia and Castellanos-Gomez, Andres and Plochocka, Paulina},
  journal={Nano letters},
  volume={18},
  number={12},
  pages={7651--7657},
  year={2018},
  publisher={ACS Publications}
}

@article{sun2023abnormal,
  title={Abnormal out-of-plane vibrational Raman mode in electrochemically intercalated multilayer MoS$_2$},
  author={Sun, Yufei and Yin, Shujia and Peng, Ruixuan and Liang, Jia and Cong, Xin and Li, Yi and Li, Chenyu and Wang, Bolun and Lin, Miao-Ling and Tan, Ping-Heng and Wan, Chunlei and Liu, Kai},
  journal={Nano Letters},
  year={2023},
 volume={23},
  number={11},
 pages={5342–5349}, 
  publisher={ACS Publications}
}

@article{ren2023autonomous,
  title={Autonomous experiments using active learning and AI},
  author={Ren, Zhichu and Ren, Zekun and Zhang, Zhen and Buonassisi, Tonio and Li, Ju},
  journal={Nature Reviews Materials},
  volume={8},
  number={9},
  pages={563--564},
  year={2023},
  publisher={Nature Publishing Group UK London}
}

@misc{Xie2024Twist2DNet,
  author = {Xie, Y.},
  title = {Twist2DNet},
  year = {2024},
  howpublished = {\url{https://github.com/YongXie-ICMM/Twist2DNet.git}},
  note = {Accessed: February 10, 2024}
}

\end{document}